\documentstyle[aps]{revtex}
\begin{document}
\newcommand{\beq}{\begin{equation}}
\newcommand{\eeq}{\end{equation}}
\newcommand{\bea}{\begin{eqnarray}}
\newcommand{\eea}{\end{eqnarray}}
\def\plumin{\underline{+}}
\def\minplu{{{\stackrel{\underline{\ \ }}{+}}}}
\def\bfr{{\bf r}}
\def\bfw{{\bf w}}
\def\bfc{{\bf c}}
\def\hpsi{\hat \psi (\bfr)}
\def\hpsid{\hat \psi^\dagger (\bfr)}
\def\tpsi{\tilde \psi (\bfr)}
\def\tpsid{\tilde \psi^\dagger (\bfr)}

%\twocolumn[\hsize\textwidth\columnwidth\hsize\csname@twocolumnfalse%
%\endcsname
\draft
%\begin{title}
%\bf
\title{Effects of temperature upon the collapse of a Bose-Einstein
condensate in a gas with attractive interactions}
%\end{title}
\author{M. J. Davis and D. A. W. Hutchinson}
%\begin{instit}
\address{
Clarendon Laboratory, Department of Physics,
University of Oxford, Parks Road,
Oxford OX1 3PU
}
\author{E. Zaremba}
%\begin{instit}
\address{
Department of Physics,
Queen's University,
Kingston, Ontario, Canada K7L 3N6
}
%\end{instit}

\date{\today}

\maketitle

\begin{abstract}

We present a study of the effects of temperature upon the excitation
frequencies of a Bose-Einstein condensate formed within a dilute 
gas with a weak attractive effective interaction between the atoms.
We use the self-consistent Hartree-Fock Bogoliubov treatment within
the Popov approximation and compare our results to previous zero
temperature and Hartree-Fock calculations
The metastability of the condensate is monitored by
means of the $l=0$ excitation frequency. As the number of atoms in 
the condensate is increased, with $T$ held constant,
this frequency goes to zero, signalling
a phase transition to a dense collapsed state. The critical
number for collapse is found to decrease as a function of 
temperature, the rate of decrease being greater than that obtained
in previous Hartree-Fock calculations.

\end{abstract}

%]

Mean field theories of the Bose-Einstein condensation of 
trapped alkali vapours 
have been extremely successful both qualitatively and 
quantitatively in determining
the excitation frequencies of the condensates, especially 
at relatively low temperatures
($\leq$ 0.7 $T_c$) \cite{dodd}. These calculations have been 
based upon the Popov approximation to
the Hartree-Fock Bogoliubov (HFB) treatment,
where the anomalous average of the fluctuating field operator
is neglected \cite{keith}. In all cases the 
study has been of alkali vapours with positive s-wave scattering
lengths (i.e., repulsive effective interactions). The case of
attractive interactions ($^7$Li for example, as used in experiments
at Rice University \cite{rice}) has not been treated in this manner.
Calculations have, rather, been based on 
the zero temperature Gross-Pitaevskii
equation (GPE) \cite{ruprecht} or upon a 
Hartree-Fock variational calculation
\cite{houbiers}. 

There are two main reasons why the HFB formalism was not used in the
Hartree-Fock study referred to above. Firstly, in the case of negative 
scattering length, the HFB-Popov collective excitations of a
homogeneous system are unstable at long wavelengths. 
Houbiers and 
Stoof\cite{houbiers} therefore found it more appealing to use the 
Hartree-Fock method, which has stable excitations
at long wavelengths. In the case of the trapped gas
this does not present a problem, as one is saved from the 
infra-red limit by the finite zero-point energy of the trap. 
From an alternative viewpoint, the finite size of the condensate
eliminates very long wavelength excitations. The HFB-Popov theory is
hence quite applicable for trapped gases. 

Secondly, there is the possibility
that atoms with an attractive effective interaction can undergo a
BCS-like pairing transition
\cite{BCSnote}. This possibility is in fact included in the full theory
that Houbiers and Stoof develop. However, in their {\it numerical} 
calculations , they ignore the 
possibility of pairing and the results presented
are based on a Hartree-Fock treatment of  Bose-Einstein 
condensation alone. If one is going to
{\it assume} that their is no BCS transition, then a better description
would appear to be that of the HFB-Popov formalism. 
This is the treatment adopted in this letter.

The purpose of the present investigation is to determine the stability
of the condensate against mechanical collapse, and the effects thereon
of thermal excitations. It has been 
shown \cite{stoof}  in the homogeneous limit that
the condensate is unstable at the densities
required for BEC.
In the trap, the additional kinetic energy can stabilise the condensate
and a metastable state is possible. This state decays on a timescale which
is long compared to the lifetime of the experiment, but
only exists for condensates below a certain size. At some critical
condensate number the condensate becomes unstable and collapses. This
instability is characterised by the monopolar collective excitation
going soft~\cite{ruprecht} (viz., the excitation frequency goes to 
zero).  Various predictions for the critical number, $N_c$,
have been made at zero temperature using the GPE and at finite 
temperatures using the Hartree-Fock treatment. Here we investigate the
effects of temperature upon the collapse via the HFB-Popov approach
as described briefly below.

We make the usual decomposition of the Bose field operator into 
condensate and
noncondensate parts; $\hpsi \equiv \Phi(\bfr) + \tpsi$.
The condensate wavefunction $\Phi(\bfr)$ is then defined
within the Popov approximation by the generalised Gross-Pitaevskii 
equation (GPE)
\beq
\left [-{\nabla^2\over 2m} + V_{ext}(\bfr) + gn_0(\bfr) +
2g\tilde n(\bfr) \right ] \Phi(\bfr) = \mu \Phi(\bfr) \,.
\label{cond2}
\eeq
Here, $n_0(\bfr) \equiv |\Phi(\bfr)|^2$ and $\tilde
n(\bfr) \equiv \langle \tpsid \tpsi \rangle$ 
are the condensate and noncondensate densities respectively.
The Popov approximation\cite{griffin,popov,shi}
consists of omitting the anomalous 
correlation $\langle \tpsi \tpsi \rangle$, but keeping $\tilde
n(\bfr)$. The condensate wavefunction in Eq.(\ref{cond2}) 
is normalised to
$N_0$, the total number of particles in the condensate.
$V_{ext}(\bfr)$ is the external confining potential and $g = 4\pi
\hbar^2a/m$ is the interaction strength
determined by the $s$-wave scattering length $a$. For $^7$Li the value
of $a$ used is -27.3 Bohr radii.
The condensate eigenvalue is given by
the chemical potential $\mu$\cite{comment}.

The usual Bogoliubov transformation, $\tpsi = \sum_i
[u_i(\bfr)\hat \alpha_i - v_i^*(\bfr)\hat \alpha_i^\dagger ]$,
to the new Bose operators $\hat \alpha_i$ and $\hat \alpha_i^\dagger$
leads to the  coupled HFB-Popov equations\cite{griffin}
\bea
\hat {\cal L} u_i(\bfr) - g n_0(\bfr) v_i(\bfr) &=& E_i
u_i(\bfr)
\nonumber \\
\hat {\cal L} v_i(\bfr) - g n_0(\bfr) u_i(\bfr) &=& -E_i
v_i(\bfr)\,,
\label{HFB}
\eea
with $\hat {\cal L} 
\equiv -\nabla^2/2m + V_{ext}(\bfr)+ 2gn(\bfr) - \mu
\equiv \hat h_0 + gn_0(\bfr)$. These
equations define the quasiparticle excitation energies $E_i$ and
the quasiparticle amplitudes $u_i$ and $v_i$. 
Once these quantities have been determined, the 
noncondensate density is obtained from the
expression\cite{griffin}
\bea
\tilde n(\bfr) &=& \sum_i \left \{ |v_i(\bfr)|^2 +
\left [ |u_i(\bfr)|^2+|v_i(\bfr)|^2 \right ] N_0(E_i)
\right \}
\nonumber 
\\ &\equiv& \tilde n_1(\bfr)+\tilde n_2(\bfr)\,,
\label{tilden}
\eea
where $\tilde n_1(\bfr)$ is that part of the density which 
reduces to the quantum depletion of the condensate as
$T \to 0$.  The component $\tilde n_2(\bfr)$ depends
upon the Bose distribution, $N_0(E) = (e^{\beta E}-1)^{-1}$,
and vanishes in the $T \to 0$ limit.

Rather than solving the coupled equations in Eq.(\ref{HFB})
directly, we introduce the
auxiliary functions
$\psi_i^{(\pm)}(\bfr) \equiv u_i(\bfr) \pm v_i(\bfr)$ which are
solutions of a pair of uncoupled equations (a more detailed
discussion of the method is presented in Hutchinson {\it et al.} of
Ref.[1]).  The two functions are related to each other by 
$\hat h_0 \psi_i^{(+)} = E_i \psi_i^{(-)}$.
We note that the collective modes of the condensate 
can be shown to have an
associated density fluctuation given by $\delta n_i(\bfr) 
\propto \Phi(\bfr) \psi_i^{(-)}(\bfr)$.

To solve these equations we introduce the
normalised eigenfunction basis defined as the solutions of
$\hat h_0 \phi_\alpha(\bfr) = \varepsilon_\alpha
\phi_\alpha(\bfr)$ and diagonalize the resulting matrix
problem. 
The lowest energy solution gives the condensate wavefunction 
$\Phi(\bfr) = \sqrt{N_0}\phi_0(\bfr)$ with eigenvalue
$\varepsilon_0 = 0$. 

The calculational procedure can be
summarised for an arbitrary confining potential as follows:
Eq.(\ref{cond2}) is first solved self-consistently for
$\Phi(\bfr)$, with $\tilde n(\bfr)$ set to zero.
Once $\Phi(\bfr)$ is known, the eigenfunctions of $\hat h_0$
required in the expansion of the excited state amplitudes are
generated numerically.  The matrix problem 
is then set up to obtain the eigenvalues $E_i$, and the corresponding 
eigenvectors $c_\alpha^{(i)}$ are used to evaluate the 
noncondensate density.
This result is inserted into Eq.(\ref{cond2}) and the
process is iterated, keeping the condensate number $N_0$ and
temperature $T$ fixed. The level of convergence is monitored by means
of the noncondensate number, $\tilde N$ and the iterations are
terminated once $\tilde N$ is within one part in $10^7$ of its
value on the previous iteration \cite{itnote}. In this  way, we generate
the self-consistent densities, $n_0$ and $\tilde n$, as a function of
$N_0$ and $T$.

We consider first the case of $T=0$ for an isotropic harmonic trap
with a frequency equal to the geometric average of the
frequencies corresponding to the Rice trap \cite{rice},
$\bar \nu=144.6$ Hz. This is the 
geometry considered previously by Houbiers and Stoof \cite{houbiers}
and with whom we find qualitatively agreement. There are several
signatures of a collapse of the condensate with increasing condensate
number $N_0$. First, we can look at the 
behaviour of the convergence parameter (the total number of particles 
in the noncondensate for a given condensate number and temperature) 
used to monitor the convergence of the solution to the HFB-Popov 
equations. In Fig. 1 we show $\tilde N$ as a function of iteration 
number for the three values $N_0=1243$, 1244, and 1245. The
convergence is clear in the first two cases, whereas in the final case
the algorithm diverges catastrophically and no stable solution can be
found. We therefore identify the critical number, $N_c$, of atoms in
the condensate as 1244, beyond which the condensate is no longer 
metastable, but unstable to the formation of a dense solid phase. 
This value of $N_c$ is slightly greater than the value of 1241
obtain by Houbiers and Stoof using the Hartree-Fock approximation.
A second, more physical indicator of the collapse is the observed 
strong dependence of the excitation frequencies on the number of 
condensate atoms. In particular, we find that the $l=0$ mode goes 
soft as $N_0$ approaches the critical number found above. We shall
focus on this criterion for the instability in the following.

We next consider a trap with confining 
frequency 150 Hz. The excitation frequencies are again calculated 
as a function of the number of particles in the condensate, both at
$T=0$ and at finite temperature. The lowest lying modes at temperatures
of 0, 200, and 400 nK are shown in Fig. 2. The lowest mode is the
$l=1$ Kohn mode, which corresponds to a rigid centre of mass motion.
For a harmonic trap the excitation frequency of this mode should be 
identically equal to the trap frequency. However, the dynamics of the
noncondensate are neglected in this treatment and the calculated 
excitations are those of the condensate alone, moving in the 
effective {\it static} potential $V_{eff}=V_{ext}+2 g \tilde n(\bfr)$.
Due to the presence of the noncondensate,
the effective potential is not parabolic and hence the generalised
Kohn theorem does not apply. The Kohn theorem is approximately obeyed
for low temperatures and low particle numbers since the noncondensate
is either small, or relatively uniform over the extent of the condensate
and hence does not introduce a significant anharmonicity. It is only
for higher temperatures near $N_c$ where the noncondensate density is
both large and sharply peaked around the centre of the trap that 
there is a marked deviation from the trap frequency. 

As mentioned above, the softening of the $l=0$ breathing mode is
a signature of the instability from a metastable condensate to a 
completely collapsed state. For $T=0$ the critical number, $N_c$, 
is found to be 1227, which is slightly lower than that obtained in the
previous case with a stronger confining potential. This is the
change in the critical number expected~\cite{ruprecht} on the basis 
of the dependence $N_c \propto 1/\sqrt{\omega_0}$, which shows that the
critical number increases as the trap confinement is relaxed.
With increasing temperature, the frequency of the $l=0$ mode is found
to go to zero at lower condensate numbers. This is because the 
attractive nature of the interactions with the thermal cloud 
creates an effective potential for the condensate which is stiffer 
than the applied external potential~\cite{houbiers}. 
The peak density of the condensate hence increases with
temperature (for fixed $N_0$) and the critical 
condensate number is reduced. The critical number for 200 nK, as obtained
from the failure to find a 
converged solution at larger $N_0$,
is $N_c=1093$. That for $T=400$ nK is $N_c=1016$.
The temperature dependence of the critical condensate number as a 
function of temperature is shown in the inset of Fig. 2.

It should be noted that the total number of trapped atoms, $N$, 
varies for each point
in the figure. Alternatively one could vary $T$ (and hence $N_0$) keeping 
$N$ fixed, which would give a critical temperature for collapse.
Experimentally evaporative cooling removes atoms. A
certain total number, corresponding to the transition temperature, 
is reached at which condensation occurs. Further cooling (removal of atoms)
then proceeds to a point where a second critical temperature (or total
number) is reached, at which point the second phase transition (i.e. collapse)
is observed. However for the experiments on $^7$Li the difference between the
critical temperatures for BEC and collapse is extremely small. Cooling thus
results in repeated collapse and growth, reducing $N$ until a stable $N_0$
is reached \cite{collapse}.

The collapse occurs because as one increases $N_0$, the peak noncondensate
density increases due to the interactions. This in turn creates a tighter 
and tighter effective potential for the condensate, which eventually results
in collapse. Fig. 3 shows the noncondensate density at 100 nK for a range
of  $N_0$. The dotted curve is for $N_0=50$, the dashed-dot-dot curve that
for $N_0=1000$. Note that for a large change in $N_0$ the peak noncondensate
density has changed relatively little. The next three curves are for $N_0=$
1100, 1130, and 1146, the final figure being the critical number. The peak
noncondensate density increases rapidly over this range. This is a cooperative
effect; the tighter effective potential reduces the frequency of the lowest
collective mode. This leads to a growth in the population
of this low lying mode, which has a
density localised near the centre of the condensate, increasing the
peak noncondensate density.

The variation of the critical number with temperature is shown in the inset
to Fig. 2. The rate of decrease of $N_c$ with $T$ is
significantly greater in the HFB-Popov treatment than in the
Hartree-Fock treatment (see Fig. 8 of Ref. 5). This is due to the different
excitation spectra calculated in the two formalisms. In our treatment we
calculate (and populate) the collective excitations, which include the 
low lying $l=0$ mode. Near collapse this mode has a much lower frequency
than the lowest single particle excitation of the Hartree-Fock spectrum. The
population of excited states is therefore underestimated in the Hartree-Fock
treatment, and as a result, the thermal population of the state is lower
than it is with
the HFB-Popov spectrum. The noncondensate population, and
hence peak density, increases more rapidly as a function of temperature in our
calculation (c.f. inset to Fig. 3 and Fig. 7 of Ref. 5). This is what gives
rise to the more rapid reduction in the critical number as the temperature is
increased.

In conclusion, we have presented the first self-consistent HFB-Popov
calculations for a dilute gas of atoms with attractive effective interactions.
We have studied the collective mode frequencies of such a gas and using these
frequencies, investigated the phase transition from metastable Bose-Einstein
condensate to a collapsed dense phase. The results from these calculations
are in general agreement with previous Hartree-Fock results, but we feel
that the HFB-Popov approach is the more appropriate one to use
if only
a BEC transition is assumed to take 
place. We find a significantly greater dependence
of the critical number upon temperature in the HFB-Popov treatment.

If one includes the possibility of a
BCS-like pairing transition then this is not the appropriate approach as
the omitted pair correlations (the so called anomalous average) are very
important. Indeed the pair correlation term, $\langle \tpsi \tpsi \rangle$,
becomes the order parameter for the BCS-like transition. The possibility of
such a transition, or the existence of mixed phases containing both BEC and
BCS macroscopic quantum states is currently under investigation.

This work was supported by grants from the Natural Sciences and
Engineering Research Council of Canada and from the United Kingdom
Engineering and Physical Sciences Research Council. We would like
to thank Henk Stoof and Keith Burnett for many helpful and enlightening
discussions.

%{

\begin{figure}
\caption{Noncondensate number as a function of the numerical
iteration number for a trap corresponding to the Rice experiment.
For $N_0=1243$ and $N_0=1244$, where $N_0$ is the number of atoms in
the condensate, a converged solution is obtained with a 
self-consistent value for $\tilde N$ which varies by less than
one part in $10^8$ between iterations. For $N_0=1245$ no stable
solution can be found. The critical number is identified as
$N_c=1244$ which is the condensate number for which the $l=0$
mode frequency goes to zero.
}
\end{figure}

\begin{figure}
\caption{Low lying mode frequencies as a function of condensate 
number for $T=0$ (solid), $T=200$ nK (dashed), and $T=400$ nK
(dotted) for a trap with a confining frequency of 150 Hz. Note
the softening of the $l=0$ mode for large $N_0$ and the decrease
in the critical number at which the mode goes soft as the 
temperature is increased. The variation of the critical number
as a function of temperature is shown in the inset.
}
\end{figure}

\begin{figure}
\caption{Noncondensate density for a range of $N_0$ at a temperature
of 100 nK in a spherical harmonic trap of frequency 150 Hz. The curves, 
in increasing order of peak density, are for
$N_0=$ 50, 1000, 1100, 1130, and 1146. The final figure is the critical
number. The inset shows the peak noncondensate density as a function
of temperature at the critical number.
}
\end{figure}


\begin{references}
\bibitem{BCSnote}Strictly the transition is an Evans-Rashid phase
transition. We will still refer to the pairing transition, however,
as a BCS or BCS-like transition as we feel this conveys the nature
of the phenomenon more clearly.
\bibitem{itnote} For calculations at high temperatures the high
lying states are simply harmonic oscillator states, which are populated 
according to the Bose distribution. The convergence criterion is based 
solely on the self-consistent states calculated from Eq.(3).
\end{references}
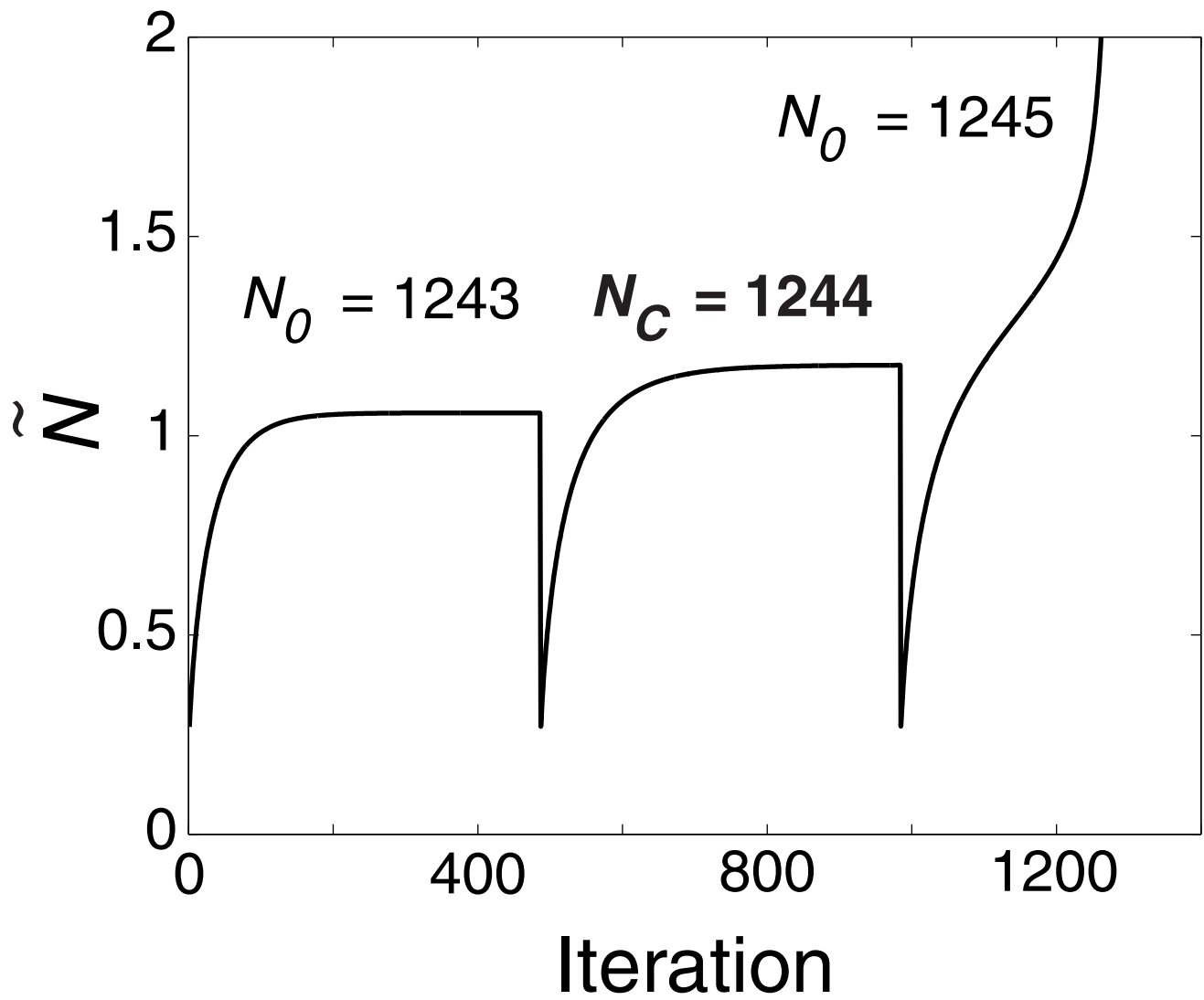
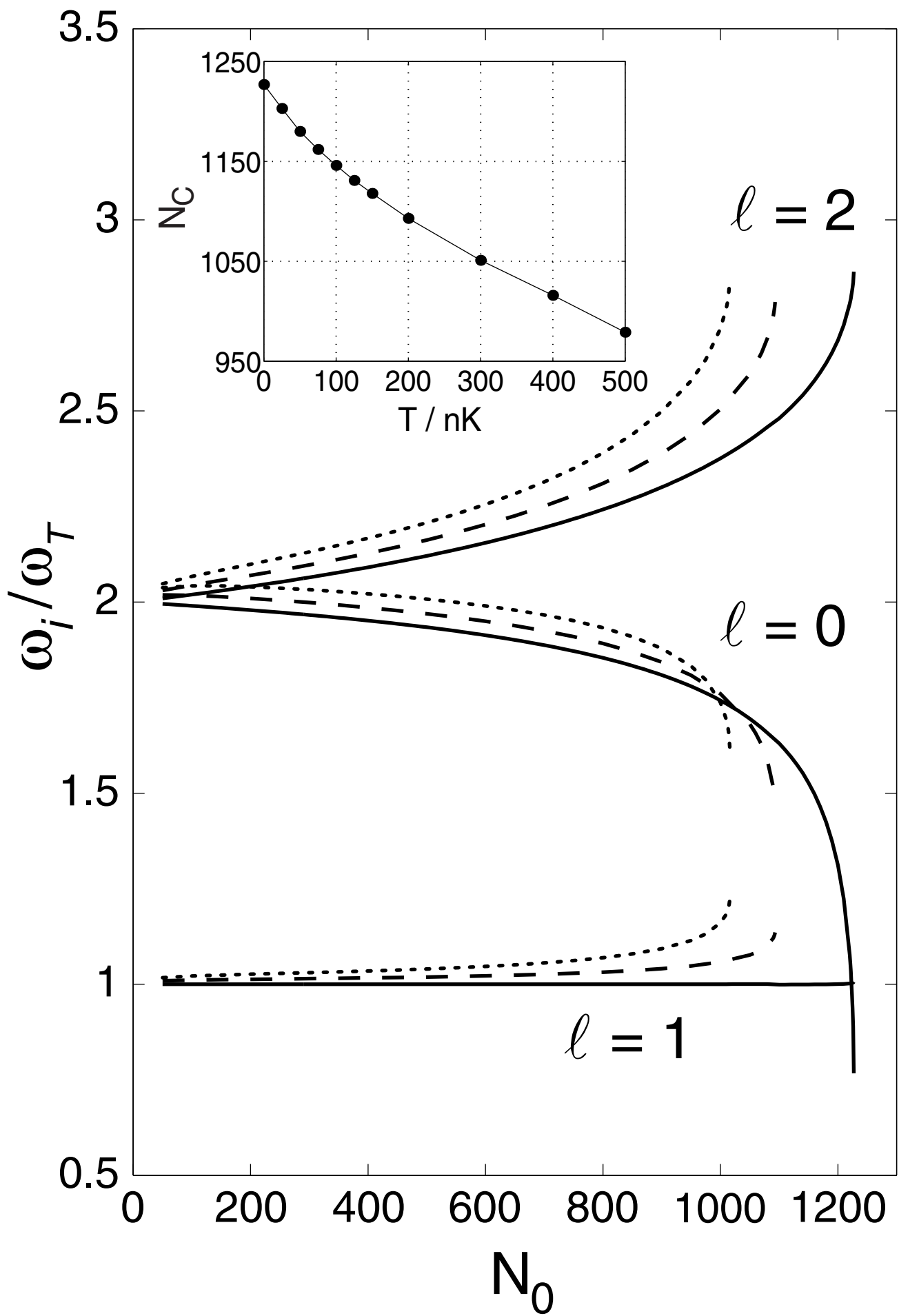
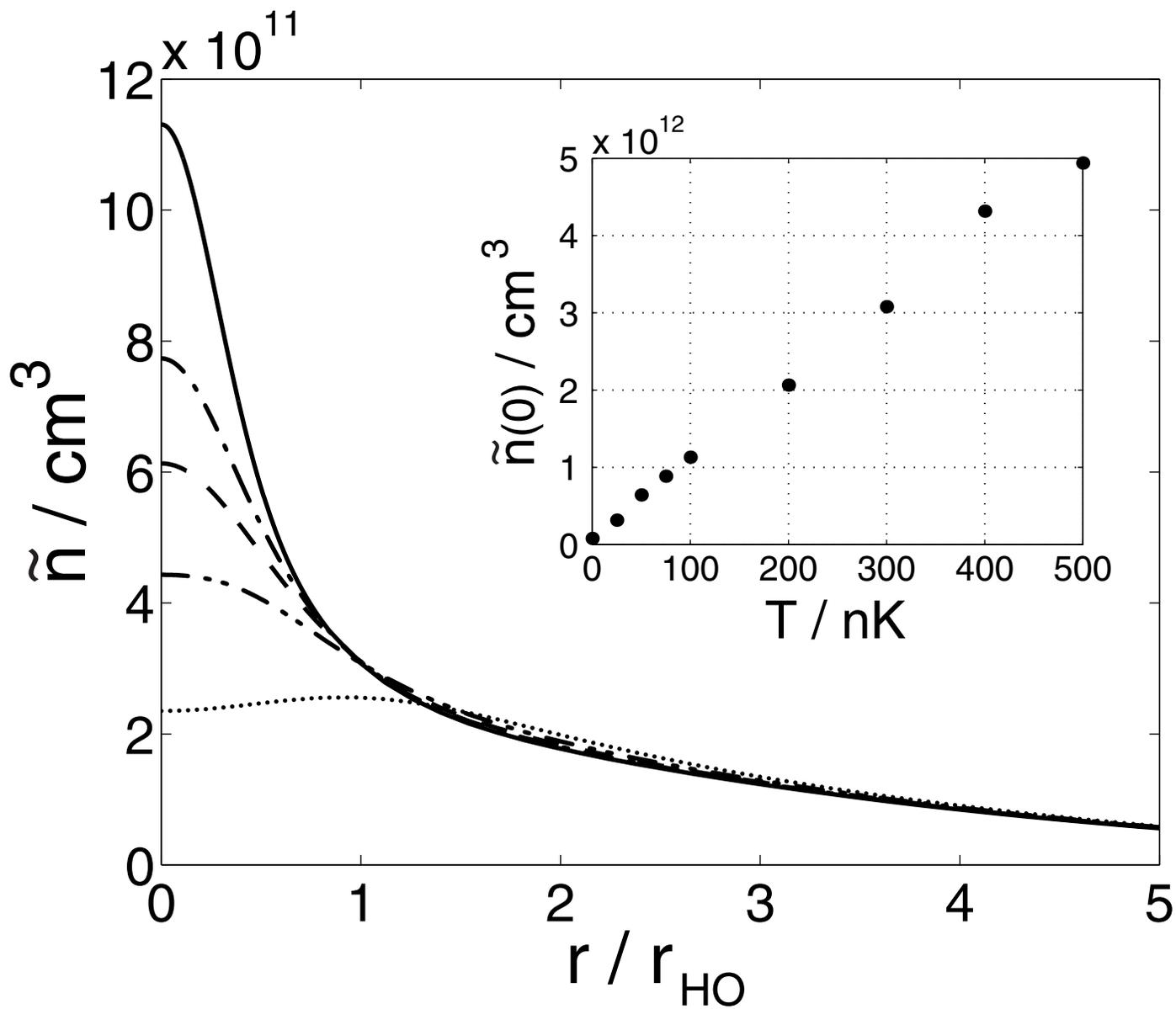
\end{document}